
%
%
%
\tolerance=100000
\magnification=1200
\baselineskip=20pt
\font\title=cmbx10 scaled \magstep2

\centerline{\title{A variational study of directed polymers}}
\centerline{\title{in disordered media with short-range correlations}}

\vskip 0.5truein

\centerline{Thomas Blum}

\vskip 0.1truein

\centerline{Department of Theoretical Physics}
\centerline{University of Manchester}
\centerline{Manchester M13 9PL}
\centerline{United Kingdom}

\vskip 0.5truein

\centerline{ABSTRACT}

\noindent A variational method that allows for replica-symmetry
breaking is applied to directed polymers in an $(N+1)$-dimensional
disordered medium.
\ The noise studied here has gaussian correlations, {\it i.e.}
it is short-ranged.
\ In dimensions $N<2$, the variational scheme yields only a
strong-coupling phase and anomalous diffusion; while in
dimensions $N>2$ it shows weak- and strong-coupling phases
but no anomalous diffusion.
\ Comparisons are made with the results of M\'ezard and
Parisi [11] for noise with power-law correlations.

\vfill

\noindent PACS: 05.40.+j, 05.20.-y,75.10.Nr

\eject

The sluggish rate of progress on strongly disordered systems attests
to the difficulty of the problem.
\ Some headway has been made over the past several years in the
area of manifolds subject to a random potential.
\ These may represent the most tractable models in which disorder
plays a crucial role.
\ Furthermore, they have been tied to various physical phenomena.
\ For instance, the one-dimensional incarnation, a directed
polymer in a random environment [1-12], has been related to domain
walls in two-dimensional disordered ferromagnets [1], surface
growth [13], and randomly stirred fluids [14].

By a ``directed" polymer, one means a polymer or random walker
that always proceeds in a positive direction along a given
coordinate in an ($N+1$)-dimensional space.
\ The effect of the disorder is manifested by the walker's
accumulating a series of random weights, one associated with each
site through which it passes.
\ The walker's attempt to maximize the benefit of the random
potential results in far greater stretching  along the axes
transverse to the directed axis than the purely entropic spreading
of its non-random counterpart.
\ In the non-random case, the transverse fluctuations are diffusive,
{\it i.e.} scale as: \ $\langle \omega^2(\ell)\rangle \sim \ell$
(where $\ell$ measures the distance along the directed axis and
$\omega$ that along the transverse axes); whereas in the random case,
the transverse fluctuations are superdiffusive, {\it i.e.} scale as
$\overline{\langle \omega^2(\ell)\rangle}\sim \ell^{2 \nu}$
(for large $\ell$) where $\nu \geq {1 \over 2}$ (where
$\overline{f(x)}$ indicates averaging $f(x)$ over realizations
of the disorder).

This nontrivial behavior is deeply rooted in the quenched nature
of the disorder --- as is the difficulty in the analysis thereof.
\ Thus, it seems natural that the issues and techniques familiar
from another problem featuring quenched randomness, namely
spin glasses, have resurfaced here.
\ Derrida and Spohn [6] made the connection to spin glasses
explicit when they uncovered a mapping between directed polymers
on a Cayley tree and the Random Energy Model, which is in some
sense ``the simplest spin glass." [14]

The replica approach, another technique from the study of spin
glasses, has made its way into the study of manifolds in
disordered media.
\ Early on, Kardar [4] used it to suggest that $\overline{\langle
\omega^2(\ell) \rangle} \sim \ell^{4 \over 3}$ for ($1+1$)-dimensional
directed polymers with delta-function correlated disorder.
\ More recently, M\'ezard and Parisi (MP) [11] applied a variational
version of the replica method to the general problem of random manifolds
pinned by quenched random impurities.
\ This variational approach has also found applications in the study
of (non-directed) polymers in a random media [16], protein-folding [17],
disordered vortex lattices [18], the random-field Ising model [19], and
interference effects in variable-range hopping [20,21].

In their extensive treatment of manifolds [11], MP included the case
of directed polymers with delta-function correlations by invoking
dimensional analysis, that is, by considering power-law correlations
that scale in the same way as the delta functions: \ for instance,
$1/|x|$ would replace $\delta^{(1)}(x)$.
\ In the present work, we apply the variational replica
method to directed polymers with noise correlations that fall
off like a gaussian $\sim {\rm e}^{-\omega^2}$ --- a genuine
short-range behavior.

To recap briefly, MP [11] studied a very generic hamiltonian for
$D$-dimensional manifolds in an $(N+D)$-dimensional space:
$$H[\omega] \ =\ {1 \over 2} \int dx \sum_{\mu=1}^D
\left( {\partial {\bf \omega} \over \partial x_{\mu}}\right)^2
+ \int dx ~V(x, {\bf \omega}(x)), \eqno(1)$$
where ${\bf \omega}(x)$ is an $N$-component vector field and $x$ is a
$D$-dimensional vector.
\ The potential $V(x,{\bf \omega})$ is a random variable with zero
mean and correlation:
$$\overline{V(x,{\bf \omega})V(x^{\prime},{\bf \omega^{\prime}})} \ =\
-N \delta^{(D)}(x-x^{\prime}) f\left( {[{\bf \omega} - {\bf
\omega^{\prime}}]^2
\over N} \right). \eqno(2)$$
The scaling with $N$ was introduced to facilitate the large-$N$ limit
in which the variational approach used is expected to become exact; it
will be maintained here for purposes of comparison.

MP considered noise correlations of the power-law form:
$$f_{p.\ell .}\left({\omega^2 \over N}\right) \ \sim \
{g \over 2(1-\gamma)}
\left( {\omega^2 \over N}
\right)^{1-\gamma}, \eqno(3) $$
for ${\bf \omega}^2 \gg 1$.
\ Under these conditions, their calculations split quite naturally into
two cases: \ noise with ``long-range" correlation (either $D\geq 2$ or
$D<2$ and $\gamma<{2 \over 2-D}$) and noise with short-range
correlation ($D<2$ and $\gamma>{2 \over 2-D}$).
\ Again, we note that MP only considered directed polymers with
delta-function correlations indirectly by relating them to
power-law correlations with the same scaling, in particular, by
setting $\gamma=1+N/2$.
\ One might find it somewhat dissatisfying that this prescription
places directed polymers with delta-function correlations in
$(1+1)$ and $(2+1)$ dimensions into their ``long-range" category.
\ Another possible source of discontentment is MP's treatment of
``short-range" case, in which they expanded the short-range behavior
of the correlation in a power series, truncating it after only two
terms.
\ The present calculation employs a bona fide short-range form for
the correlation and requires no such truncation.
\ It will be shown to reproduce the main results of ref. [11]
for directed polymers.

In the directed-polymer case ($D=1$), the partition function
corresponding to $H[\omega]$ (eq. (1)) is a path integral,
$$Z(\omega,x) \ =\ \int_{\{0,0\}}^{\{\omega,x\}}
{\cal D} \omega ~ {\rm exp}\Bigl\{
-\beta H[\omega] \Bigr\} \eqno(4a)$$
(where $\beta$ is the reciprocal temperature) which can be shown to
obey the equation:
$${\partial Z(\omega,x) \over \partial x} \ =\ -{\cal H}(\omega)
{}~Z(\omega,x), \eqno(4b)$$
where ${\cal H}$ is the (random) operator:
$$  {\cal H}(\omega) \ =\ -~{1 \over 2 \beta} \sum_{i=1}^N
{\partial^2 \over \partial \omega_{i}^2} ~+~
\beta V(x,{\bf \omega(x)}). \eqno(4c)$$
But it can be usual first to replicate the system $n$ times and average
over the disorder; one then obtains a path integral which obeys
$${\partial \overline{Z^n(\{\omega_a\},x))} \over \partial x}
\ =\ -{\cal H}_n(\{\omega_a\}) ~\overline{Z^n(\{\omega_a\},x)},
\eqno(5a)$$
with the following (non-random) $n$-body hamiltonian:
$$  {\cal H}_n(\{\omega_a\}) \ =\ -~{1 \over 2 \beta} \sum_{i=1}^N
\sum_{a=1}^n {\partial^2 \over \partial
\omega_{ai}^2} ~+~ {\beta^2 \over 2} \sum_{a \neq
b}^n N f \left({1 \over N } ({\bf \omega_{a}} -
{\bf \omega_{b}})^2 \right), \eqno(5b)$$
with attractive interactions given by $f$ (as defined in eq. (2)).

The interesting exact solutions to such an $n$-body
Schr\"odinger-like equation have already been exploited, [4,8]
and standard perturbative approaches have been shown to yield
to unphysical results. [11]
\ Hence, we shall proceed here with a variational scheme of the
Rayleigh-Ritz type.
\ That is, we shall study the ground-state energy of
${\cal H}_n(\{\omega_a\})$ using the following variational wave
function:
$$\Psi \ =\ {\cal N} \ {\rm exp} \ \Bigl\{ -{1 \over 2} \sum_{a,b}^n
\sum_{i=1}^N \hat m_{a,b} \ \omega_{i a}\ \omega_{i b} \Bigr\}.
\eqno(6)$$
This method of searching for the variational ground state of an
operator hamiltonian is convenient when considering one-dimensional
manifolds ($D=1$) [12,17,20,21] and differs slightly from the approach
taken in ref. [11] where higher dimensional manifolds are considered
as well.
\ The results, however, do not differ.

Keeping in mind the eventual $n \rightarrow 0$ limit and the
possible emergence of replica-symmetry breaking (RSB) therein, a
gaussian form was chosen for calculational ease prior to that
limit.
\ The matrix $\hat m$ of variational parameters is chosen as an
$n \times n$ hierarchical matrix with $K$-step breaking (as
introduced by Parisi [22]).
\ In the one-step ($K=1$) breaking scenario, one can envision
breaking the replicas into groups with the couplings between
``particles" belonging to the same group differing from the
couplings between ``particles" belonging to different groups.
\ Next, two-step breaking divides the groups into subgroups, and
so on, leading to the hierarchical (ultrametric) structure.
\ Eventually, the limits $K \rightarrow \infty$ and $n \rightarrow
0$ are to be taken.
\ In addition to offering one more variational parameters (than
the replica-symmetric version), it is hoped that the variational
scheme with RSB can better mimic situations in which two or more
diverging paths have nearly degenerate energies, including those
with the very lowest energies. [9]

With the gaussian wave function, calculating  $\langle {\cal H}_n
\rangle_{\Psi}$ requires computing $Tr(\hat m)$ (which arises from
the expectation of the kinetic-energy portion of ${\cal H}_n$)
and certain functions of $(m_{a,a}^{-1} +m_{b,b}^{-1}-2m_{a,b}^{-1})$
(which emerge from the interaction terms).
\ To be more specific, a Taylor expansion and Wick's theorem applied
to an interaction term lead to [11]:
$$ \eqalignno{
 \Biggl \langle f\biggl[ {({\bf \omega_{a}} - { \bf \omega_{b}})^2
\over N }\biggr]\Biggr \rangle_{\Psi}
\ =\ &\sum_j {f^{(j)} \over j! N^j} \Bigl\langle
\bigl( {\bf \omega_{a}} - {\bf \omega_{b}}\bigr)^{2j}
\Bigr\rangle_{\Psi} \cr
\ =\ &\sum_j { f^{(j)} \over j! N^j} \
{(N+2j-2)!! \over (N-2)!!} \
{\Bigl( m^{-1}_{a,a} + m^{-1}_{b,b} -2
m^{-1}_{a,b}\Bigr)^j \over 2^j}\cr
\ =\ &\hat f \biggl[ {m^{-1}_{a,a} + m^{-1}_{b,b} -2
m^{-1}_{a,b} \over 2 } \biggr], &(7a) \cr}$$
where
$$\hat f(z) \ =\ {1 \over \Gamma({N \over 2})} \int_{0}^{\infty}
d \alpha \ \alpha^{{N \over 2}-1} {\rm e}^{-\alpha} f\left(
{2 \alpha z \over N}\right). \eqno(7b)$$
We are currently interested in the case in which the correlations
have a gaussian form:
$$f_{g} \left({ ({\bf \omega_{a}} - {\bf \omega_{b}})^2
\over N} \right)
\ =\ -\left({\mu^2 \over \pi N}\right)^{N \over 2}
{\rm exp}\left\{{-\mu^2 \over N } ({\bf \omega_{a}} -
{\bf \omega_{b}})^2 \right\}, \eqno(8a)$$
where $N^{1 \over 2}/\mu$ serves as a correlation length and
$f_g(z)$ is normalized here so that it approaches a delta function in the
$\mu \rightarrow \infty$ limit.
\ The corresponding $\hat f$ is:
$$\hat f_g(z) \ =\ -\left({\mu^2 \over \pi N}\right)^{N \over 2}
\left[ 1 + {2 \mu^2 z \over N}\right]^{-{N \over 2}}. \eqno(8b)$$

For power-law correlations (eq. (3)), the same procedure yields:
$$\hat f_{p.\ell .}(z) \ = \
{g~\Gamma \Bigl( 1 + {\textstyle{N \over
2}} -\gamma \Bigr) \over 2~(1-\gamma) ~ \Gamma \Bigl(
{\textstyle{N \over 2}} \Bigr) }
\Biggl( {2z \over N} \Biggr)^{1-\gamma},
\eqno(9)$$
provided $\gamma \leq 1 + {N \over 2}$.
\ Note that $\hat f_g(z)$ and $\hat f_{p. \ell.}(z)$ have the
same large-$z$ behavior when $\gamma=1 + {N \over 2}$, furnishing
evidence for the validity of MP's dimensional-analysis approach
within this scheme.
\ (Setting $\gamma=1+{N \over 2}$ requires some small-$z$
regularization; otherwise, the coefficient of $\hat f_{p. \ell.}$
would diverge.)

The $K$-step hierarchical matrix $\hat m$ has $K+2$ parameters
$\{\tilde a,a_0,a_1,\ldots,a_K\}$ which take on the form $[\tilde a,
a(u)]$, where $a(u)$ is a function on the interval $[n,1]$, in
the $n\rightarrow 0$ limit. [22]
\ Ref. [21] showed how the eigenvalues of $\hat m$, which
take on a similar form $[\tilde \lambda, \lambda(u)]$, are related
to $[\tilde a, a(u)]$.
\ In terms of the eigenvalues, $\langle {\cal H}_n \rangle_{\Psi}$
becomes:
$$
{ \langle {\cal H} \rangle_{\Psi} \over Nn} \ =\
{1 \over 4\beta}\int_1^n {du \over u^2} \lambda(u)
{}~+~{1 \over 4\beta n}  \tilde \lambda ~+~
\beta^2  \int_1^n du \hat f \bigl[ Q(u)\bigr],  \eqno(10a)$$
where
$$ Q(u) \ =\ \int_1^u {dv \over v^2}
\lambda^{-1}(v) ~+~ {1 \over u}\lambda^{-1}(u). \eqno(10b)$$
The first two terms in eq. (10a) correspond to Tr($\hat m$),
{\it i.e.} the sum over the eigenvalues.
\ Note that $du/u^2$ is roughly the degeneracy of $\lambda(u)$.

Now comes the variation.
\ One obtains solutions for the ``best" $\lambda(u)$ through the
following procedure.
\ First, take a functional derivative of eq. (10a) with respect to
$\lambda(u)$ and set it equal to zero; this yields the stationarity
equation:
$$\lambda^2(u) \ =\ 4 \beta^3 \Biggl\{
\int_u^n dv \hat f^{\prime} \bigl[ Q(v) \bigr]
+u\hat f^{\prime}\bigr[ Q(u) \bigl] \Biggr\},\eqno(11)$$
where $\hat f^{\prime}=\partial \hat f/\partial Q$.
\ As a step toward finding an equation which is local in $u$
({\it i.e.} no integrals over $u$), take a derivative with respect
to $u$, which results in:
$$2 \lambda(u) \lambda_u(u) \ =\ -4 \beta^3
\hat f^{\prime \prime}[Q(u)]\lambda^{-2}(u)\lambda_u(u),
\eqno(12a)$$
where $\lambda_u=\partial \lambda/\partial u$.
\ This result implies that either $\lambda_u(u)=0$ or
$$2^{-1}\beta^{-3} \lambda^3(u)  \ =\ -
\hat f^{\prime \prime}[Q(u)].\eqno(12b)$$
To pursue the latter case, put in the desired form of
$f^{\prime \prime}(x)$:
$$\hat f_g^{\prime \prime}(x) \ =\ - {\mu^{4} (N+2)
\over  N} \left( {\mu^2 \over \pi N} \right)^{N \over 2}
\left[ 1 + {2 \mu^2 x \over N}
\right]^{- {N+4 \over 2}} \eqno(13a)$$
and for simplicity let
$$s \ =\ {\mu^{4}(N+2) \over N}
\left( {\mu^2 \over \pi N} \right)^{N \over 2} \eqno(13b)$$
and
$$t \ =\ {2 \mu^2 \over N}. \eqno(13c)$$
These substitutions produce:
$$ 2^{-1}\beta^{-3} \lambda^3(u)  \ =\
s \Bigl[1+tQ(u)\Bigr]^{-{N+4 \over 2}} \eqno(14a)$$
which upon inversion becomes:
$$ \Bigl[ 2^{-1}\beta^{-3} s^{-1} \lambda^3(u)
\Bigr]^{-{2 \over N+4}} \ =\ 1 +tQ(u), \eqno(14b)$$
where $Q(u)$, recall, is given by eq. (10b).
\ Now, eq. (14b) is still non-local in $u$, so take another
derivative with respect to $u$, furnishing:
$$-~\left({6 \over N+4} \right)
\bigl(2 \beta^3 s \bigr)^{{2 \over N+4}}
{}~\lambda^{-{N+10 \over N+4}}(u)
{}~\lambda_u(u) \ =\ - ~{t \over  u}
\lambda^{-2}(u)~\lambda_u(u). \eqno(15)$$
The solution of eq. (15) is either $\lambda_u(u)=0$ or:
$$\lambda(u) \ = \
\bigl( 2 \beta^3 s \bigr)^{{2 \over 2-N}}
\left({6 u \over (N+4)t}
\right)^{{N+4 \over 2-N}} . \eqno(16)$$
Note that eq. (16) and $\lambda(u)=Const.$ (which corresponds
to $\lambda_u(u)=0$) are merely the possible local
solutions.
\ One must piece together a function on the entire interval
$[n,1]$, consisting of these local solutions, which satisfies
the (non-local) stationarity equation (eq. (11)).

The pattern of RSB found by MP in their  ``long-range" solution
fuses together a power-law behavior (similar to eq. (16))
for $n\leq u<u_c$ and a constant for $u_c<u\leq 1$.
\ MP demonstrated that this small-$u$, power-law dependence is
intimately connected to the property of superdiffusion: \
$\overline{\langle {\bf \omega}^2(\ell)\rangle} \sim
\ell^{2 \nu}$ with $\nu>{1 \over 2}$.
\ In particular, they found for $\nu$ the ``Flory" exponents [5]
({\it e.g.} $\nu = {3 \over 5}$ for directed polymers in
$1+1$ dimensions, instead of the exact result $\nu={2 \over 3}$
[3,4]).

For the gaussian problem, we will search for solutions of a similar
nature.
\ Note, however, that for $N>2$, the power-law is not well behaved
for small $u$ in the $n \rightarrow 0$ limit --- then  we will
seek solutions like those derived by MP for their short-range case;
which consist of two constants ({\it i.e.} one-step breaking).
\ Hence, the present calculation also splits into two parts:
\ $N<2$ where one finds the full-breaking, power-law solution for small
$u$ and which displays anomalous diffusion; and $N>2$ where one finds the
one-step breaking solution, a phase transition (see below),
but no anomalous diffusion.

First, for the case $N<2$, let us assume a solution of the form:
$$\lambda(u) \ =\ \cases{
\bigl( 2 \beta^3 s \bigr)^{{2 \over 2-N}}
\biggl[ {6 u \over (N+4)t} \biggr]^{{N+4 \over 2-N}}
, & \ \ if \ $n<u<u_c$; \cr
\bigl( 2 \beta^3 s \bigr)^{{2 \over 2-N}}
\biggl[ {6 u_c \over (N+4)t} \biggr]^{{N+4 \over 2-N}}
, & \ \ if \ $u_c<u<1$; \cr} \eqno(17)$$
then one can determine $u_c$ by considering eq. (14) at $u=1$
(bypassing some tedious integration).
\ This procedure yields the following equation for $u_c$:
$${N \over \mu^2} \left[ { N (N+2)
\beta^3  \over 4 \pi^{{N \over 2}} }\right]^{{2 \over 2-N}}
\left({6u_c \over N+4} \right)^{{6 \over 2-N}}
{}~+~ {6u_c \over (N+4)}
{}~-~1 \ =\ 0. \eqno(18)$$
This solution, eqs. (17) and (18), was also obtained numerically
by solving iteratively a discretized version of the stationarity
equation (eq. (11)), giving credence to the assumptions made in
(17).

Note that here $u_c$ is $\beta$-dependent.
\ This feature differs from MP's solution for power-law correlations,
where $u_c$ is a constant. [11]
\ With a $\beta$-dependent $u_c$, one ought to check for
solutions with $u_c$ lying outside the interval $[0,1]$,
as that might indicate some kind of a phase transition, but $u_c$
always lies within this range for $N<2$.
\ When $\beta = 0$ or $\mu = \infty$ (the delta-function limit),
the first term in eq. (18) is zero, and one finds $u_c =
{N+4 \over 6}$, which is precisely the constant found by MP for
power-law correlations (if $\gamma=1+N/2$).
\ Then, as $\beta$ increases (or $\mu$ decreases)  $u_c$
approaches though does not reach zero, so that $0 < u <{N+4 \over 6}$.
\ For finite $\mu$ and large $\beta$, one finds $u_c \sim
\beta^{-1}$.
\ This property distinguishes predictions concerning the
large $\beta$ behavior of quantities (such as the free energy)
made by the gaussian and power-law calculations.

For $N>2$, we look for a solution with one-step breaking of the
form:
$$\lambda(u) \ =\ \cases{ 0, & if $0<u<u_c$; \cr
                          A, & if $u_c<u<0$. \cr} \eqno(19)$$
Substituting this solution into the hamiltonian (eq. (10a))
yields:
$${\langle {\cal H} \rangle_{\Psi} \over Nn} \ =\
{1 \over 4 \beta} \left( 1 - {1 \over u_c} \right)A ~+~
\beta^2  (1-u_c) \left({ \mu^2 \over \pi N}
\right)^{N \over 2} \left(1 + {t \over A } \right)^{-{N \over 2}} .
\eqno(20)$$
Note that when $\lambda(u)=0$, $Q(u)=\infty$.
\ Next, taking derivatives with respect to $A$ and $u_c$ and setting
them equal to zero results in:
$$u_c \ =\ { A^2 \over 2 N  \beta^3 t }
\left({\pi N \over \mu^2}\right)^{N \over 2}
\left( 1 + {t \over A} \right)^{{N +2\over 2}}
\eqno(21)$$
and
$$u_c^2 \ =\ {A \over 4 \beta^3 }
\left({\pi N \over \mu^2}\right)^{N \over 2}
\left( 1 + {t \over  A} \right)^{N \over 2},
\eqno(22)$$
the stationarity equations.

Eliminating $A$ leads to:
$$1 \ =\ {t \over 8 \beta^3 u_c^3} \left(
{\pi N^2 \over \mu^2} \right)^{N \over 2}
(N-2u_c)^{1-N/2}
\eqno(23)$$
Again, we check for situations in which $u_c$ lies outside the
interval $[n,1]$.
\ This time we find some.
\ In fact, we find $\beta_c$, the critical $\beta$, simply by
setting $u_c=1$ in eq. (23):
$$ \beta_c \ =\ {t^{1 \over 3} \over 2}
\left( {\pi N^2 \over \mu^2} \right)^{N \over 6}
\Bigl( N-2 \Bigr)^{{2-N \over 6}}
. \eqno(24)$$
For high temperatures, $\beta < \beta_c$, the solution of eq. (23)
has $u_c>1$; so then the solution is $\lambda(u)=0$ for the entire
interval $[n,1]$ --- the (trivial) replica-symmetric solution.
\ Here too, the assumptions made in eq. (19) were tested using the
discretized version of eq. (11).

The presence of a transition to a weak-coupling phase for $N>2$
is in accord with the known behavior of directed polymers. [6]
\ As mentioned earlier, in their consideration of short-range
correlations, MP expanded $\hat f(z) \approx f_0 + f_1 z + \ldots$
for small $z$, keeping only the first two terms in their explicit
calculations.
\ In the gaussian case, no such approximation was deemed
necessary, and yet the same basic results were obtained.
\ Hence, it indicates that MP's results are not purely an
artifact of the truncation procedure.

In summary, we have applied a variational replica method to
$(N+1)$-dimensional directed polymers with gaussian-correlated
noise.
\ The calculation divided quite naturally into two regimes.
\ In the first, $N<2$, the variational solution displayed
full replica-symmetry breaking with a power-law behavior for
small $u$.
\ This feature is associated with superdiffusion, and in particular
the ``Flory" exponents for the transverse fluctuations are found.
\ Similar behavior was observed by MP [11] in their
so-called ``long-range" solution for noise with power-law correlations.
\ Another trait shared by the gaussian-noise and power-law-noise
calculations [11] is the lack of a transition to a weak-coupling
(high-temperature) phase for $N<2$.
\ On the other hand, one distinction between the two is their
dependence on the inverse temperature $\beta$, especially for large
$\beta$.
\ In the second regime, $N>2$, the variational solution displayed
a phase with no replica-symmetry breaking for high temperatures and
a phase with one-step replica-symmetry breaking at lower
temperatures, but no anomalous diffusion.

The variational scheme manages to uncover the phase transition where
it is expected ($N>2$) and the absence of a transition where it is
expected ($N<2$); that is, it obtains the correct phase diagram.
\ It predicts anomalous diffusion for $N<2$; however, it fails to
find the correct exponent $\nu$ for the transverse fluctuations.
\ Moreover, it predicts no superdiffusion in the strong-coupling
phase for $N>2$ as is anticipated from the simulations. [10]
\ One scheme for furthering calculations involves an expansion
(in $1/N$) around the solution at $N=\infty$ which is proposed
to be exact. [23,24]
\ Another conceivable (albeit less controlled) approach would be
to improve the gaussian assumption for the variational wave
function, perhaps by using a Lanczos-type approach [24] to go
beyond the basic Rayleigh-Ritz method.
\ It is hoped that the present work has resolved a few of the
nagging issues found in MP.
\ However, the more important issues remain unresolved and more
work is required before this chapter can be closed.

\vskip 0.25truein

I would like to acknowledge useful discussions with
Yonathan Shapir and Yadin Y. Goldschmidt and to thank the
Science and Engineering Research Council, UK for financial
support.

\vfill
\eject

\noindent References

\vskip 0.25truein

\item{[1]} Huse D A and Henley C L, Phys. Rev. Lett.
{\bf 54} (1985) 2708.

\item{[2]} Kardar M and Zhang Y-C, Phys. Rev. Lett.
{\bf 58} (1987) 2087.

\item{[3]} Huse D A, Henley C L and Fisher D S, Phys. Rev. Lett.
{\bf 55} (1985) 2924.

\item{[4]} Kardar M, Nucl. Phys. {\bf B 290} (1987) 582.

\item{[5]} Nattermann T, J. Phys. {\bf C 18} (1985) 6661;
Kardar M, J. Appl. Phys. {\bf 61} (1987) 3601.

\item{[6]} Derrida B and Spohn H, J. Stat. Phys. {\bf 51} (1988)
817.

\item{[7]} Medina E, Hwa T, Kardar M and Zhang Y-C,
Phys. Rev. {\bf A 39} (1989) 3053;
Imbrie J Z and Spencer T, J. Stat. Phys. {\bf 52} (1988)
609.

\item{[8]} Parisi G, Rend. Acad. Naz. Lincei {\bf XI-1}
(1990) 3.

\item{[9]} M\'ezard M, J. Phys. France {\bf 51} (1990) 1831.

\item{[10]} Kim J M, Moore M A and Bray A J, Phys. Rev.
{\bf A 44} (1991) 2345.

\item{[11]} M\'ezard M and Parisi G, J. Phys. I (France) {\bf 1}
(1991) 809 and J Phys. {\bf A 23} (1990) L1299.

\item{[12]} M\'ezard M and Parisi G, J. Phys. {\bf A 25}
(1992) 4521.

\item{[13]} Kardar M, Parisi G and Zhang Y-C, Phys. Rev. Lett.
{\bf 56} (1986) 889.

\item{[14]} Forster D, Nelson D and Stephen M, Phys. Rev {\bf A 16}
(1977) 732.

\item{[15]} Derrida B, Phys. Rev. {\bf B 24} (1981) 2613; Gross D J
and M\'ezard M. Nuc. Phys. {\bf B 240} (1984) 431.

\item{[16]} Edwards S F and Muthukumar M, J. Chem Phys. {\bf 89}
(1988) 2435.

\item{[17]} Shakhnovich E I and Gutin A M, J. Phys. {\bf A 22}
(1989) 1647.

\item{[18]} Bouchaud J-P, M\'ezard M and Yedidia J S, Phys. Rev.
Lett. {\bf 67} (1991) 3840 and Phys. Rev. {\bf B 66} (1992)
16686; Giamarchi T and Le Doussal P {\it Elastic theory of
pinned flux lattices} (1993) preprint.

\item{[19]} M\'ezard M and Young A P, Europhys. Lett. {\bf 18}
(1992) 653.

\item{[20]} Goldschmidt Y Y and Blum T, J. Phys. I (France) {\bf 2}
(1992) 1607.

\item{[21]} Blum T and Goldschmidt Y Y, J. Phys {\bf A 25}
(1992) 6517.

\item{[22]} M\'ezard M, Parisi G and Virasoro M A, {\bf Spin
Glass Theory and Beyond} (World Scientific, Singapore 1987).

\item{[23]} Goldschmidt Y Y, Nucl. Phys. {\bf B 393} (1993)
507; {\it Manifolds in random media: \ Beyond the variational
approximation} (1993) preprint.

\item{[24]} Cook J and Derrida, Europhys. Lett. {\bf 10} (1988) 195;
J. Phys. {\bf A 23} (1990) 1523.

\item{[25]} Dagotto E and Moreo A, Phys. Rev. {\bf D 31} (1985)
865.

\vfill
\eject

\end
\bye